\documentstyle[12pt,moriond]{article}
\begin{document}
\heading{DUST IN STAR-FORMING GALAXIES}

\author{Daniela Calzetti $^{1}$} {$^{1}$ Space Telescope Science Institute, 
3700 San Martin Drive, Baltimore, MD 21218, U.S.A.} 

\begin{moriondabstract}
I review the effects of dust obscuration in galaxies at both low and
high redshifts, and briefly discuss a method to remove dust reddening
from the emerging light of star-forming galaxies. I also analyze the
evolution of the dust opacity in galaxies as a function of redshift,
and discuss its effect on the observed UV-optical light. The
quantitative corrections for dust obscuration given here allow one to
recover the intrinsic value of the global star formation at different
cosmological times.
\end{moriondabstract}

\def\putplot#1#2#3#4#5#6#7{\begin{centering} \leavevmode
\vbox to#2{\rule{0pt}{#2}}
\includegraphics{#1}
\end{centering}}
%

\section{Are Galaxies Opaque?}

The general question is: how much of the light produced by all
galaxies is absorbed by dust? Answering this question is crucial for
understanding the evolution and the star formation history of
galaxies.  It is also an intrinsically difficult task, because: (1)
dust is a source of continuous opacity and lacks readily recognizable
signatures, such as absorption or emission features, in the
UV-optical-nearIR regime; (2) the complex geometrical distribution of
dust and stars affects both the global attenuation and the reddening
of the emerging light from unresolved galaxies; (3) the colors
produced by dust reddening and by age are often degenerate.

The geometry of the dust is the dominant factor which determines the
appearance of a galaxy (Witt, Thronson, \& Capuano 1992, Calzetti,
Kinney \& Storchi-Bergmann 1994, Witt \& Gordon 1996). This is
especially true when the galaxy is unresolved, and only its spatially
integrated light can be measured. Large amounts of dust can be hidden
by a clumpy distribution; if the dust clumps are small and compact,
the interclump regions provide enough clear lines of sight that the
galaxy will appear almost dust-free. An homogeneous mixture of dust
and gas has a similar effect: the major contribution to the emerging
light comes from the outermost layers of the galaxy, while the inner
regions are generally opaque; the emerging spectral energy
distribution then appears almost unreddened, but much dimmer than in
the dust-free case. The total mass and stellar content of the galaxy
will thus be underestimated by an amount which depends in a nonlinear
fashion on the total dust content.

The absence of recognizable features has one exception: the 2200~\AA~
``bump'', namely the broad UV absorption feature characteristic of the
dust extinction in our Galaxy. The ``bump'', however, cannot be used
as a reliable gauge of the amount of dust in a galaxy. Its intrinsic
depth is different in different galaxies, with the LMC extinction
curve in the 30 Doradus region showing a shallower bump than the
average Galactic extinction curve and the SMC extinction not showing a
bump at all. The LMC and and SMC are the only two external galaxies
where accurate measures of interstellar reddening have been possible
(see, however, Bianchi et al. 1996 for the case of M31). The bump even
changes characteristics along different lines of sight within the
Milky Way itself (Cardelli, Clayton \& Mathis 1989). The nature of the
variability of the bump from galaxy to galaxy is not entirely clear,
although it probably reflects variable grain size distributions and/or
changing ratios in the dust ingredients as the characteristics of the
local ISM and environment vary (see Gordon \& Clayton 1998). Finally,
even if the extinction curves were independent of the galaxy
characteristics, the observed depth of the 2200~\AA~ bump is a
function not only of the total amount of dust along the line of sight,
but also of the geometrical distribution of the dust relative to the
emitters (Natta \& Panagia 1984).

A dust reddened stellar population may resemble an ``old''
population. The dust-age degeneracy is an important effect when only
broad band colors or low resolution spectroscopy are available for the
object of interest and stellar features are not identifiable. In
numbers: an optical attenuation A$_V$=1 changes the colors enough to
mimic an age ``increase'' of a factor of $\sim$5 in a stellar population
less than 100~Myr old. Conversely, the colors of a relatively old
stellar population will not be easily discriminated from the colors of
a young but reddened population.
 
\subsection{Opacity at Low Redshift}

Given the difficulties listed above, different authors have taken
different approaches to the problem of determining the opacity of 
galaxies in the local Universe.

The cleanest technique for measuring the dust opacity of a galaxy is
the one which measures the obscuration produced by a foreground galaxy
on background sources. The background sources can be either distant
galaxies (Gonzalez et al. 1998) or another nearby galaxy overlapping
with the first along the line of sight (e.g.: White, Keel \& Conselice
1996, Berlind et al. 1997). With this technique, the light source is
external to, and thus uncontaminated by, the galaxy for which the
opacity is to be measured. Therefore, all complications due to the
complex geometry of the dust inside the galaxy are circumvented, as
the foreground galaxy acts as a dust screen in front of the background
sources. To date, studies have concentrated on spiral galaxies as
foreground objects. These have been shown to have transparent interarm
regions, with A$_B\sim$A$_I\sim$0.1-0.2, and opaque arms, with
A$_B\sim$A$_I\sim$1. As expected, the opaque regions of spirals
coincide with the regions of most intense star formation (the arms),
as stars form in the dusty molecular clouds. Using the variation of
the galaxy surface brightness with inclination, Giovanelli et
al. (1995) have shown that the same level of opacity as the arms is
found in the centers of the spirals.

Another potentially powerful technique to determine the opacity of a
galaxy is to measure the ratio between the UV-optical-nearIR stellar
emission and the far-infrared (FIR) dust emission. This ratio
represents the amount of stellar radiation which directly escapes the
galaxy versus the amount of stellar radiation which has been absorbed
by dust and is re-rediated in the FIR. The power of the method is that
it relies uniquely on energy conservation. The shortcoming is that an
accurate determination of the energy balance requires the sampling of
the entire wavelength range from the far-UV to the far-IR, while, in
general, only a few data points are available along the spectrum. A
wealth of data exist for galaxies at optical, near-IR and, thanks to
IRAS, in the 8-120~$\mu$m region, but only relatively recently UV
measurements and FIR data beyond 120$\mu$m are becoming available for
large enough samples of galaxies. UV data are crucial for an accurate
energy balance because the UV can be energetically dominant if recent
star formation is present in the galaxy. Over the last couple of
years, ISO has added data points in the 120-240~$\mu$m regime,
covering another crucial wavelength regime: the one which contains the
peak of dust emission from quiescent galaxies. Studies employing UV,
optical, and IRAS data have concluded that in disk galaxies the amount
of stellar light absorbed by dust and re-emitted in the FIR range is
about the same as the emerging stellar light (Soifer \& Neugebauer
1991, Xu \& Buat 1995, Wang \& Heckman 1996). We have learned during
this Conference that preliminary results from ISO observations
indicate that about 30--50\%~ of the UV emission from the CFRS
galaxies is hidden by dust (Hammer 1998, these Proceedings).

Although none of the above results can be considered final, there is
mounting evidence that galaxies in the local Universe are generally
not very opaque. The available data currently suggest that dust absorbs
$\sim$1/2, and probably no more than 2/3, of the total stellar light
in local galaxies.

\subsection{Opacity at High Redshift}

At high redshift, the sampled wavelength baseline is more limited than
at low redshift, as the rest-frame UV and optical range fall in the
observed optical and IR range; thus determinations of opacities are
even less secure than in the local Universe. One could expect dust
opacity to be a minor issue at high redshift, as young galaxies were
also metal-poor and, therefore, dust-poor. However, low metallicities
were coupled with large gas column densities, which can in principle
produce non negligible opacities. In addition, standard observing
techniques target the rest-frame UV emission from high redshift
galaxies, a wavelength range very sensitive to the obscuring effects
of dust.

Meurer et al. (1997) have been among the first to raise the issue of
dust opacity for the galaxy population at z$\sim$3, by noticing that
the UV spectral energy distributions of the Lyman-break galaxies
(Steidel et al. 1996) are too red to be dust-free star-forming
objects. They attempt a correction for dust reddening of the high
redshift galaxies by extrapolating a method used for local starburst
galaxies (Calzetti et al.  1994, Calzetti 1997), and conclude that on average
$\sim$9/10 of the UV light is lost to dust obscuration.
Rowan-Robinson et al. (1997) use ISO to detect the FIR emission from
distant galaxies in the Hubble Deep Field and conclude that about 4/5
of the UV light at z$>$2 is absorbed by dust. Near-IR spectroscopy of
six of the Lyman-break galaxies has yielded measurements of the
nebular emission lines in the rest-frame optical, which indicate that
between 1/3 and 1/6 of the UV light escapes from the galaxies (Pettini
et al. 1998; Pettini 1998, these Proceedings).

\section{Is Dust Opacity Relevant for Dwarf Galaxies?}

Dust formation is strictly linked to metal production. Thus, low
metallicity systems, such as dwarf galaxies, are expected to contain
little dust. Although observations confirm this to be a generally true
statement, the global effect of the dust on the emerging light greatly
depends on its geometrical distribution within the galaxy, as stated 
above.

The case of NGC5253 is enlightening in this respect. This dwarf
galaxy, located in the Centaurus Group $\sim$4~Mpc away from the Milky
Way (Sandage et al. 1994), has a metallicity
$\sim$0.1~Z$_{\odot}$. HST images show NGC5253 as a bright UV emitter
(e.g. Meurer et al. 1995, Calzetti et al. 1997), a consequence of the
fact that the central region of the galaxy is undergoing a powerful
burst of star formation. IRAS and other measurements, however,
indicate that the galaxy's central region is also a bright
far-infrared emitter, with a FIR luminosity comparable to the B-band
emission of the entire galaxy. The FIR emission is a clear indicator
of the presence of dust, and, in this case, of dust heated by the
recently formed stars (Aitken et al. 1982). A detailed analysis of the
reddening characteristics of the starburst in NGC5253 shows that about
2/3 of the galaxy UV (2,600~\AA) light is absorbed by dust (Calzetti
et al. 1997). In conclusion, despite the fact that the galaxy is
metal-poor and, therefore, should contain a small amount of dust, a
non negligible fraction of its light is still lost to dust
absorption. The galaxy is neverthless bright in the UV, a sign that
the inhomogeneous distribution of dust produces a `picket-fence'
geometry, with `holes' in the dust distribution through which the UV
light can emerge.

\section{Reddening in Actively Star-Forming Galaxies}

The starburst regions of galaxies are characterized by high energy
densities (e.g. Kennicutt 1989); massive star winds and supernova
explosions inject energy into the ISM creating shock waves and gas
outflows.  Shocks from supernovae are most likely responsible for the
destruction of dust grains, via grain-grain collisions and sputtering
(Jones et al. 1994).  Gas outflows can develop into ``superwinds''
(Heckman, Armus \& Miley 1990) and eject significant amounts of the
interstellar gas and dust from the site of star formation. As a
result, the starburst environment is likely to be rather inhospitable
to dust (Calzetti, Kinney \& Storchi-Bergmann 1996). 

If little diffuse dust is present within the starburst site, the main
source of opacity is given by the dust {\it surrounding} the
site. This dust is internal to the galaxy, but external (or mostly
external) to the starburst region. This leads to a simplified
description of the distribution of the dust affecting the starburst
population: the geometry is equivalent to a dust shell surrounding a
central light source. Such description is rather accurate for the
reddening affecting the optical-nearIR nebular gas emission (Calzetti
et al. 1996). Figure~1a plots the color excess E$_g$(B$-$V) measured
from two pairs of nebular line ratios, the widely used
H$\alpha$/H$\beta$ and the long-baseline H$\beta$/Br$\gamma$, for a
sample of starbursts. The two color excesses are sensitive diagnostics
of dust geometry because of their difference in wavelength
baseline. The data points of Figure~1a lay along the lines which
identifies the homogeneous or clumpy screen dust models, while the
other configuration, i.e. the homogeneous mixture of gas and dust, is
excluded. An immediate consequence is that the color excess
E$_g$(B$-$V) measured from {\it any} pair of hydrogen recombination
lines in the wavelength range (0.48--2.2~$\mu$m) is a reliable
indicator of dust reddening for the gas in the starburst. It should be
remarked that the screen model is an approximated description of the
reddening affecting the nebular lines. For instance, dark and compact
dust clumps within the starburst site are not excluded by the data.

\begin{figure}[h]
\putplot{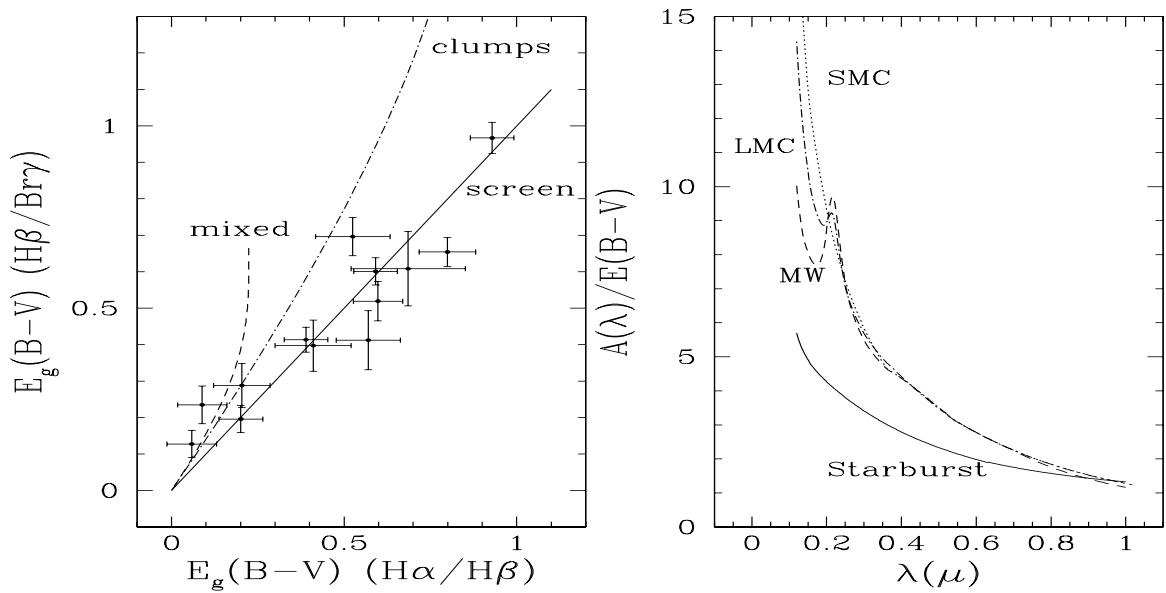}{7.8 cm}{-90}{60}{40}{-10}{230}
\caption{\small {\bf (a)} (left): The color excess E$_g$(B$-$V)
derived from the hydrogen line ratio H$\beta$/Br$\gamma$ as a function
of the same quantity derived from the ratio H$\alpha$/H$\beta$. The
error bars are 1~$\sigma$ uncertainties. The data are compared with
three models for the dust distribution: a homogeneous foreground {\it
screen}; a {\it clumpy} foreground layer with an average of $\cal
N_{clumps}$=10 along the line of sight; and a homogeneous {\it
mixture} of dust and gas, which simulates the case of internal
dust. For small E$_g$(B$-$V), i.e. small reddening values, the models
are degenerate; however, large E$_g$(B$-$V) can be explained only by
the foreground distributions (homogeneous and clumpy). Clumpy dust geometries
with more than 10 clumps along the line of sight lay between the
continuous and dot-dashed lines. {\bf (b)} (right): The {\it
Starburst} reddening curve, derived from
continuum-color~vs.~gas-reddening plots like those in Figure~2, is
shown as a function of wavelength, and compared with the interstellar
extinction curves of the Milky Way, of the 30~Dor region in the LMC,
and of the SMC.}
\end{figure}

If the {\it observed} stellar continuum colors are plotted as a
function of E$_g$(B$-$V), the two quantities correlate over the entire
wavelength range 0.12-2.2~$\mu$m (Figure~2). This correlation spans a
much wider range of colors than expected from variations of the
intrinsic stellar populations in the starburst (see, e.g., the models
by Leitherer \& Heckman 1995). The most straightforward interpretation
of the data in Figure~2 is that the stars are affected by a change in
dust reddening proportionally to the way the ionized gas is
affected. The reddening measured from the gas can then be used as the
independent parameter to derive a ``reddening curve'' for the stellar
continuum of starbursts (Calzetti et al. 1994; Calzetti 1997). The
reddening curve is shown in Figure~1b and compared with the
interstellar extinction curves of the Milky Way, the LMC and the
SMC. However, the reddening curve is {\it not} an extinction curve. It
is derived from the spatially integrated colors of the entire stellar
population in the starburst; it represents the `net' reddening, which
includes the effects of both dust geometry and composition. It mostly
describes dust absorption, since the effects of scattering are
averaged out by the fact that we are observing the entire starburst,
and therefore scattering out of the line of sight is compensated by
scattering into the line of sight.  In this light, the comparison of
the starburst reddening curve with the extinction curves of Figure~1b
is purely illustrative.

Two characteristics of the reddening curve are immediately apparent in
Figure~1b. First, the curve lacks the 2200~\AA~ bump of the Milky Way
or LMC extinction curves. Second, it has a much shallower UV-optical
slope than the other curves. The first characteristic has been proven
to be intrinsic to the {\it extinction} curve of starbursts (Gordon,
Calzetti \& Witt 1997). No dust geometry is able to `wash away' the
bump if it is present in the extinction curve; thus the intrinsic
extinction curve of starbursts must resemble the `bump-less' SMC
curve. The second characteristic is a combined effect of dust geometry
and age segregation of the stellar populations (Calzetti et
al. 1997). The dust in the starburst is likely to be clumpy, with the
young, ionizing stars closely associated with the dust clumps. Stars
are born in dusty molecular clouds, and the youngest stars have not
had enough time to move away from the parental cloud. Conversely, the
old, non-ionizing stars have lived long enough to migrate from their
birth site and spread across the starburst region, filling also the
dust-free interclump regions. Only the young stars ionize the gas,
while both young and older stars contribute to the UV-optical
continuum emission from the starburst. Therefore, the emission from
the dust-associated gas is more reddened than the integrated
UV-optical stellar continuum, resulting in a shallow reddening curve
for the stellar population. This picture is a somewhat simplified
description of what happens in a starburst, but has the power to
provide a physical understanding of the observational evidence.

The reddening curve discussed in this section is an effective tool for
removing the dust effects from the emerging light of starbursts. It
has been derived entirely from observations and is totally
model-independent.  In addition, it can be used in the same way as
standard extinction curves, in that the attenuation of the stellar
continuum is given by:
A$_{star}$=E$_g$(B$-$V)~k$_{starburst}$($\lambda$). It should be
stated again that this is only a convenient representation and does
not imply that the dust affecting the stellar continuum is in a
foreground homogeneous screen. The reddening curve of Figure~1b is
generally applicable to starburst regions, but not to subsonic HII
regions or to quiescent galaxies.

The effects of dust on local, UV-bright starburst galaxies can be
summarized as follows: the median attenuation at 1600~\AA~ is 1.6~mag,
and at H$\alpha$ is $\sim$0.8~mag. These numbers are important because
star formation rates derived from the {\it observed} UV flux will be
generally underestimated by a factor $\sim$5 (and this sample is
UV-selected!), while the same rates derived from the H$\alpha$ flux
will be underestimated by a factor of 2. Even worse, about 30\%~ of
the young stars in this UV-bright sample are completely buried in dust
and do not contribute either to the UV emission or to the nebular
lines (Calzetti et al. 1995).

\begin{figure}[h]
\putplot{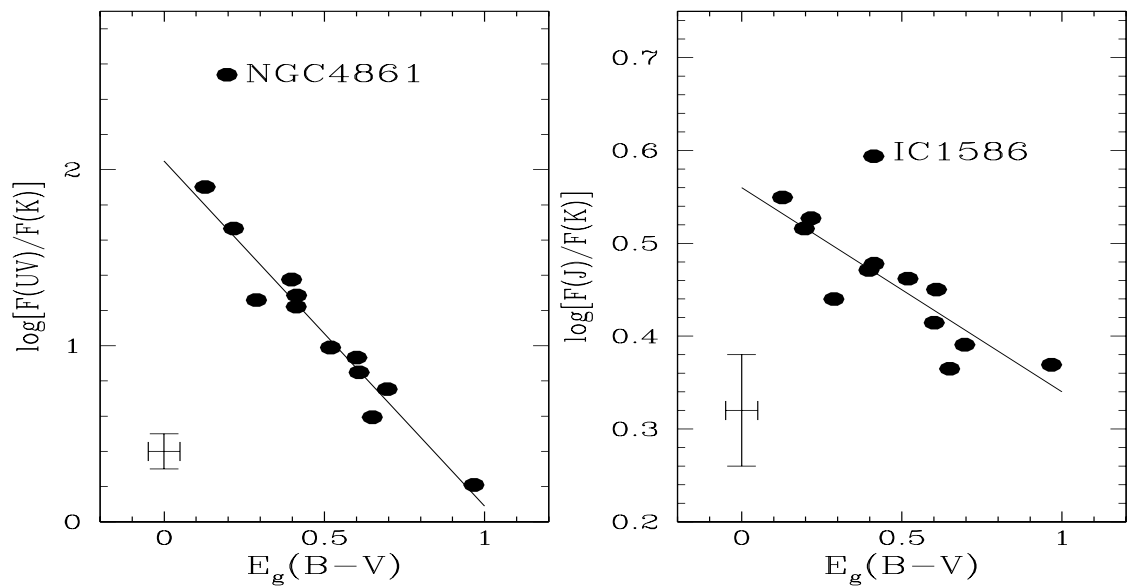}{7.8 cm}{-90}{60}{40}{-10}{230}
\caption{\small The observed UV$-$K (left panel) and J$-$K (right
panel) colors of starbursts correlate with the reddening of the
ionized gas, E$_g$(B$-$V).  The correlation indicates that the
observed continuum colors are dominated by the effects of dust
reddening, rather than variations in the intrinsic stellar
populations. These and other similar correlations have been used by
Calzetti et al. (1994) and Calzetti (1997) to derive the starburst
reddening curve of Figure~1b. The 1~$\sigma$ uncertainties on the data
points are indicated in the lower left corners of the two panels. The
starburst galaxies NGC4861 (left panel) and IC1586 (right panel)
deviate from the general trend set by the other galaxies. The
deviation of NGC4861 is due to its much younger stellar population
relative to the average; for IC1586 the culprit are the large 
photometric uncertainties (Calzetti 1997).}
\end{figure}

\section{The Evolution of Dust Opacity}

The increasing number of intermediate and high redshift galaxies which
are being secured (e.g. Lilly et al. 1996, Steidel et al. 1996) have
led to the first attempt to measure the evolution of the galaxy light
density in the Universe. An extensive work was done by Lilly et
al. (1996) who collected and used the CFRS data to derive the galaxy
luminosity density in the redshift range 0$\le$z$\le$1. Madau et
al. (1996 and 1998) and Connolly et al. (1997) extended the plot by
Lilly et al.  to the redshift range 1$\le$z$\le$4, using the HDF data
(Williams et al.  1996). Adopting an average star formation history
for the galaxies and a stellar initial mass function (IMF), Madau et
al. (1996) converted the UV light density into a global star formation
rate (SFR). A number of authors have, however, warned against the
danger of using the {\it observed} UV flux density as an indicator of
SFR, because of its sensitivity to dust obscuration (e.g. Meurer et
al. 1997, Rowan-Robinson et al. 1997). Although observations are being
obtained to try to quantify the effects of dust obscuration at high
redshift (e.g., Pettini et al. 1998), the issue is far from being a
simple one because of the potential degeneracy of UV/optical
indicators (see the discussion in Meurer, Heckman \& Calzetti 1998).

For the reasons given in Section~1.2, the low metallicity of high
redshift galaxies does not necessarily imply low dust opacities. To
test this statement, one can attempt to follow the history of the
metal and dust enrichment of galaxies using as zero-order
approximation the SFRs derived from the {\it observed} UV flux
densities as a function of redshift. Star formation implies metal and
dust production, which can be converted into opacities with an
educated guess on the dust geometry. The derived dust opacities are
then used to correct the observed UV flux densities for the effects of
dust obscuration, so that new values of the SFR can be obtained. This
procedure is then repeated until the derived UV opacities and the
ratio of observed-to-corrected UV flux densities converge to the same
value (Calzetti \& Heckman 1998). The convergence of the method is
ensured by the fact that the increase in the global SFR leads to a
faster consumptions of the available gas and, although also the metal
production increases, the general effect is to decrease the opacities,
which therefore get closer in value to the correted-to-observed UV
ratio. Outflows/inflows are used to control the metal enrichment of
the galaxies, which we impose to have a final gas metallicity
Z$_{gas}$=Z$_{\odot}$.

Other authors (e.g. Pei \& Fall 1995) have discussed the impact of
dust obscuration in models of chemical enrichment of the Universe;
this is the first time, however, that the {\it observed} SFR is used
in an iteratire procedure to derive the {\it intrinsic} SFR.

\begin{figure}[h]
\putplot{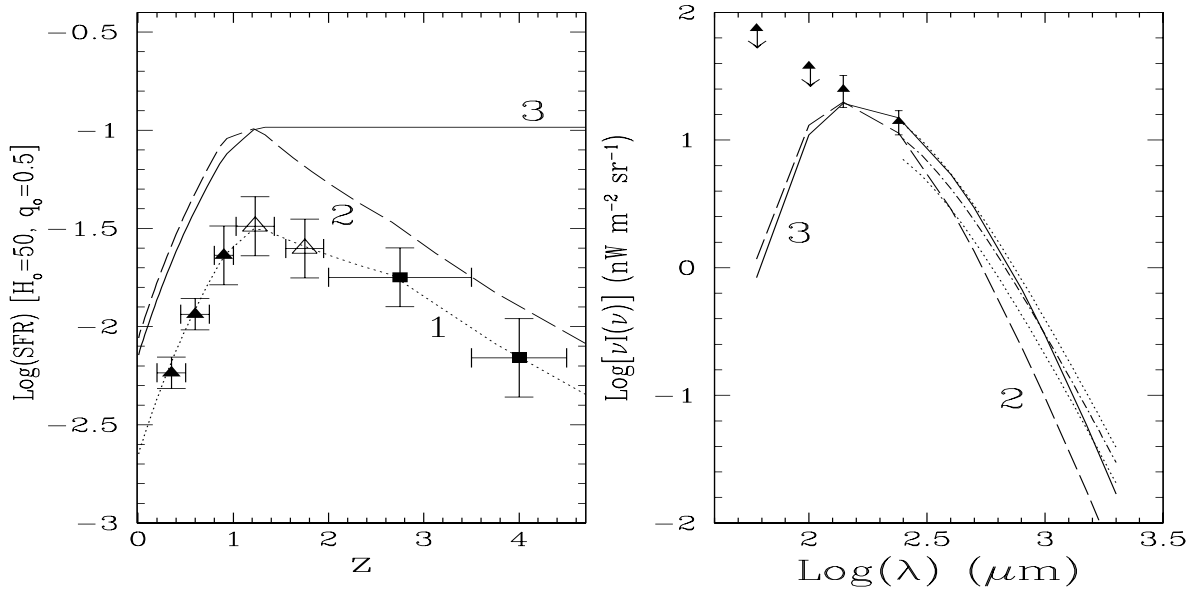}{7.8 cm}{-90}{60}{40}{-10}{230}
\caption{\small {\bf (a)} (left) The global star formation rate as a
function of redshift (Madau et al. 1996), renormalized to the adopted
stellar IMF. The data points are from the observed UV flux densities
reported by Lilly et al. (filled triangles), Connolly et al. (1997),
and Madau et al. (1998). Model~1 is a smooth representation of the
`observed' SFR. Models~2 and 3 are the solutions of the iterative
process described in the text. {\bf (b)} (right) The Cosmic Infrared
Background observed by COBE is compared with the predictions from the
Models. The DIRBE data (Hauser et al. 1998) are represented by
triangles with their 1~$\sigma$ uncertainty; the values at 60 and
100~$\mu$m are upper limits. The FIRAS data (Fixsen et al. 1998) are
shown as a smooth curve (dot-dashed line) and its 1~$\sigma$
uncertainty (dotted curves). The prediction from the Models are shown
as the continuous curve (Model~3) and the long-dashed curve
(Model~2).}
\end{figure}

Assumptions on the model are that $\sim$10\%~ of the baryons are in
galaxies; that merging is not a major galaxy formation process; that
the half light radius of the galaxies increases with time (Giavalisco
et al. 1996); that the stellar IMF is constant in time and with a
Salpeter slope in the range 0.35--100~M$_{\odot}$, with no stars
formed below 0.35~M$_{\odot}$; that instantaneous mixing holds for the
metals and dust in galaxies; that the outflows/inflows are
proportional to the SFR (Pei \& Fall 1995); that the metal/dust ratio
evolves with metallicity (Pettini et al. 1997); that the dust is
homogeneously mixed with the stars (Xu \& Buat 1995, Wang \& Heckman
1996).  The UV flux density at z=4 (Madau et al. 1998) is not included
in the iterative procedure because, to date, there are only a few
candidates which have been confirmed at that redshift, and volume
corrections are currently unknown (Dickinson 1998, private
communication). The uncertainty on the z=4 point is potentially large.

The observational constraints to the solutions are, in addition to the
observed UV flux density as a function of redshift, the local
dereddened H$\alpha$ energy density (see Gronwall 1998, these
Proceedings; Gallego 1998, these Proceedings), the local FIR density
(Soifer \& Neugebauer 1991, Villumsen \& Strauss 1987), and the Cosmic
Infrared Background from COBE (CIB; Fixsen et al. 1998, Hauser et
al. 1998).

Given the limited amount of observational constraints and the
associated uncertainties, the method converges to multiple solutions,
two of which (Figure~3a) represent extreme behaviors in the allowed
range. One of the two solutions (Model~2) reproduces relatively well all
observational constraints, except the CIB at long wavelengths
(Figure~3b). Model~2 implies a modest dust correction, a factor
$\times$2, to the UV flux densities at z$>$2 and a larger factor,
$\sim\times$4, at z$<$1. The total amount of stars produced by this
solution is modest enough that no outflows/inflows are required to
keep the final gas metallicity to Z$_{\odot}$; the solution actually
produces a final gas metallicity 3/4~Z$_{\odot}$. Model~2 has a peak
at z$\sim$1 and its shape closely resembles that of the observed SFR
(Model~1 in Figure~3a).

The other solution, Model 3 (Figure~3a), reproduces the observational
constraints, including the long wavelength regime of the CIB
(Figure~3b), better than Model~2. Modest outflows/inflows are required
for this solution, implying that galaxies were in the past a factor
$\sim$1.7/1.5 heavier/lighter, respectively, than today. The UV dust
reddening correction is $\times$3.2 at z$<$1 and greater than
$\times$5 at z$>$2. The shape of Model~3 resembles the `monolithic
collapse' scenario for star formation. In Model~3, $\sim$20\%~ of the
stars formed at z$>$3.

Although Model~3 meets the observational constraints better than
Model~2, current uncertainties, especially at high redshift, do not
allow us to accept or reject altogether either of the solutions. It
is, however, noteworthy that even in the most favorable case metal
enrichment implies a dust correction factor of at least 2, and most
likely as large as a factor of 5--6, on the observed UV flux densities
and, therefore, on the global SFR at z$>$2.

\section{Summary}

Low redshift galaxies are probably not very opaque: multiwavelength
measurements indicate that the fraction of light absorbed by dust is
about 1/2, and no more than 2/3, of the total stellar light. In the
case of actively star-forming regions, a general tool is available to
correct the observed UV/optical/near-IR emission for the effects of
dust obscuration and, thus, recover the intrinsic flux.

High redshift data are still insufficient to answer the question of
how much opaque young galaxies are. Estimates range from about 1/2 to
about 9/10 of the total stellar light lost to dust absorption.
Simulations of the evolution of the dust content and opacity of
galaxies suggest that, in the most optimistic case, about 1/2 of the
light is lost to dust absorption at z$>$2. However, the most
optimistic case may not be the most likely one, and larger values of
the opacity are possible at high redshift. Far-infrared observations
more sensitive than those provided by ISO are needed to unambiguosly
address this issue.

\vspace{0.5cm}
{\bf Acknowledgements.} The author gratefully acknowledge the Conference 
organizers and the STScI DDRF for providing financial support.

\vfill
\end{document}